\documentclass[aps,prstab,reprint,groupedaddress,amsmath,amssymb,showpacs,showkeys,floatfix]{revtex4-1}

\usepackage{graphicx}
\usepackage{dcolumn}
\usepackage{bm}

\begin{document}

\title{Notes about collision monochromatization in $e^+e^-$ colliders}

\author{ A.~Bogomyagkov }
  \email{A.V.Bogomyagkov@inp.nsk.su}
  \affiliation{Budker Institute of Nuclear Physics SB RAS, Novosibirsk 630090, Russia}
  \affiliation{Novosibirsk State University, Novosibirsk 630090, Russia}
\author{ E.~Levichev }
  \affiliation{Budker Institute of Nuclear Physics SB RAS, Novosibirsk 630090, Russia}
  \affiliation{Novosibirsk State Technical University, Novosibirsk 630073, Russia}

\date{\today}

\begin{abstract}
The manuscript describes several monochromatization schemes starting from A.~Renieri \cite{ref:Renieri} proposal for head-on collisions based on correlation between particles transverse position and energy deviation. We briefly explain initial proposal and expand it for crossing angle collisions. Then we discuss new monochromatization scheme for crossing angle collisions based on correlation between particles longitudinal position and energy deviation.
\end{abstract}


\maketitle

\section{Introduction}
\begin{figure}[h]
\includegraphics[width=\columnwidth,trim=10 20 20 15, clip]{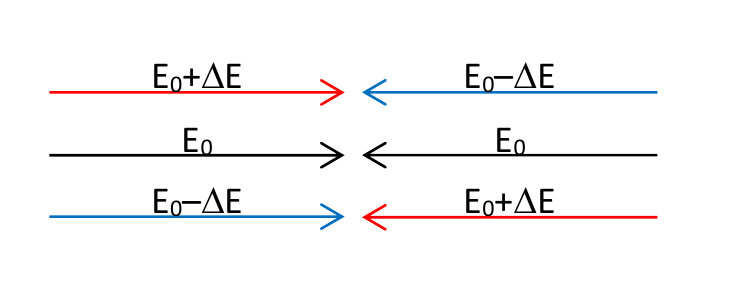}
\caption{Monochromatization scheme for head-on collisions.}
\label{fg:Monochromatization-head-on}
\end{figure}

The energy resolution of existent colliders is very large with respect to the energy width of narrow resonances. For example FCC-ee \cite{ref25,ref26} at beam energy 62.5~GeV to study Higgs boson at the threshold production provides energy resolution of 140~MeV while the width of Higgs boson is 4.2~MeV. In 1975 A.~Renieri \cite{ref:Renieri} proposed to improve energy resolution of Italian collider Adone \cite{ref:Amman:1966,ref:Bassetti:1974} by introduction of dispersion function of opposite signs for colliding beams at the interaction point (IP), thus creating a correlation between particles transverse position and energy deviation (FIG. \ref{fg:Monochromatization-head-on}).
During the following years physicists proposed to upgrade existing colliders or to build new ones based on the same monochromatization principle: VEPP-4\cite{ref:Avdienko:1988}, Tau-Charm factory\cite{ref:Zholents:1992,ref:FausGolfe:1992}, SPEAR \cite{ref:Wille:1984}, B factory\cite{ref:Dubrovin:1991}, LEP\cite{ref:Bassetti:1987}. However, the monochromatization principle was never tested. 

Recent proposal of monochromatization for FCC-ee\cite{ref:ValdiviaGarcia:2016} was a reason for us to return to the topic of monochromatization in head-on collisions, derive expressions for monochromatization with crossing angle and propose a new monochromatization scheme for crossing angle collisions based on particle's longitudinal position correlation with energy deviation.

The starting point for our calculations is definition of luminosity $\mathcal{L}$ as a ratio of number of events per second $\dot{N}$ and a total cross section $\sigma$ \cite{ref:Moller:1947,ref:Suzuki:luminosity}
\begin{equation}
\label{eq:luminosity-0}
\mathcal{L}=\frac{\dot{N}}{\sigma}=f_0(1+\cos(2\theta))\int n_1n_2dV dct\,,
\end{equation}
where $f_0$ is bunch collision rate, $n_1$ and $n_2$ are the densities of colliding bunches, $2\theta$ is the full crossing angle (FIG. \ref{fg:Geometry-collisions}), $V$ is a volume occupied by bunches, $c$ is speed of light and $t$ is time.

\section{Invariant mass}
\begin{figure}[h]
\includegraphics[width=\columnwidth,trim=0 17 0 17, clip]{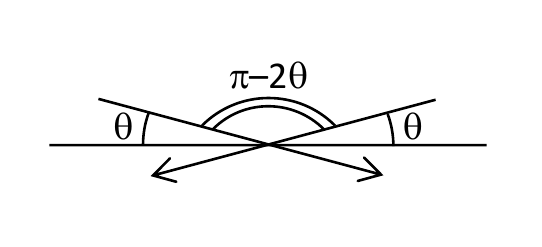}
\caption{Collision geometry}
\label{fg:Geometry-collisions}
\end{figure}

Collider energy resolution is defined as a square root of second central moment of luminosity distribution with respect to invariant mass $M$. For two colliding particles with four-momenta $P_1^\mu=\{E_1,\vec{p}_1\}$ and $P_2^\mu=\{E_2,\vec{p}_2\}$ invariant mass is
\begin{equation}
M^2=(P_1^\mu+P_2^\mu)^2=2m_e^2+2E_1E_2-2\vec{p}_1\vec{p}_2\,,
\label{eq:InvariantMass-1}
\end{equation}
where $\theta$ is a half of the crossing angle (FIG. \ref{fg:Geometry-collisions}), $E_1$, $E_2$ and $\vec{p_1}$, $\vec{p_2}$ are the energies and momenta of two colliding particles, $m_e$ is electron mass.
We will neglect electron mass in further calculations and assume the speed of light $c=1$, therefore energy and momentum are equal $E=\left|\vec{p}\right|$. To calculate the scalar product of momenta we choose accompanying first particle coordinate system with axis $z$ directed along the azimuth of beam orbit
\begin{align}
\vec{p}_1&=\{p_{1,x},p_{1,y},p_{1,z}\}\,, \\
\vec{p}_2&=\{-p_{2,x}'\cos(2\theta)+p_{2,z}'\sin(2\theta)\,, \nonumber \\
             & \qquad \; p_{2,y}'\,, \\
             & \qquad \! -p_{2,x}'\sin(2\theta)-p_{2,z}'\cos(2\theta)\}\,, \nonumber
\end{align}
where the prime variables describe coordinates of the second particle in self accompanying system.
Introducing normalized variables $\delta_{1,2}=(E_{(1,2)}-E_0)/E_0$, $x_{1,2}'=p_{(1,2),x}/p_0$ and $y_{1,2}'=p_{(1,2),y}/p_0$ we obtain for longitudinal momentum
\begin{equation}
\begin{split}
p_{(1,2),z}&=\sqrt{p_0^2(1+\delta_{(1,2)})^2-p_{(1,2),x}^2-p_{(1,2),y}^2} \\
			&=p_0\left(1+\delta_{(1,2)}-\frac{x_{(1,2)}'^2}{2}-\frac{y_{(1,2)}'^2}{2} \right)+O(3)\,,
\end{split}
\end{equation}
where $E_0$ and $p_0=E_0$ are the average energy and momentum for both bunches.
Substituting obtained expression in \eqref{eq:InvariantMass-1} we obtain
\begin{equation}
\begin{split}
M^2&=2E_0^2\biggl[(1+\delta_1+\delta_2+\delta_1\delta_2)(1+\cos(2\theta)) \\
	&+ (x_2'-x_1'+x_2'\delta_1-x_1'\delta_2)\sin(2\theta) - y1'y2'\\
	&-\left(\frac{x_1'^2+x_2'^2+y_1'^2+y_2'^2}{2}-x_1'x_2'\right)\cos(2\theta)\biggr]+O(3)\,.
\end{split}
\label{eq:InvariantMass-2}
\end{equation}
Now, we assume that colliding bunches' population obey normal distribution with average energy and momentum $E_0=\left<E_1\right>=\left<E_2\right>$ and $E_0=p_0$, and standard deviation $\sigma_E$ with relative value $\sigma_\delta=\sigma_E/E_0$, with angular spreads in vertical and horizontal planes $\sigma_{x'}=p_{x'}/p_0$ and $\sigma_{y'}=p_{y'}/p_0$. We also take a square root of \eqref{eq:InvariantMass-2} which gives us
\begin{equation}
\begin{split}
M&=2E_0\biggl[\biggl(1-\frac{y_1'^2+y_2'^2}{4}-\frac{(x_1'-x_2')^2}{8} \\
  & - \frac{(\delta_1-\delta_2)^2-4(\delta_1+\delta_2)}{8}\biggr)\cos(\theta) + \frac{(y_1'-y_2')^2}{8 \cos(\theta)}\\
  & +\frac{2(x_2'-x_1')+(\delta_1-\delta_2)(x_2'+x_1')}{4}\sin(\theta)\biggr]+O(3)\,.
\end{split}
\label{eq:InvariantMass-3}
\end{equation}

Averaging \eqref{eq:InvariantMass-2} and \eqref{eq:InvariantMass-3} over angular spreads we derive expressions ready for further use with luminosity distribution:
\begin{equation}
\begin{split}
\left<M^2\right>_{x',y'}&= 2E_0^2(1+\delta_1+\delta_2+\delta_1\delta_2)(1+\cos(2\theta)) \\
& -2E_0^2(\sigma_{x'}^2+\sigma_{y'}^2)\cos(2\theta)+O(3)\,, 
\end{split}
\label{eq:InvariantMass-4}
\end{equation}
\begin{equation}
\begin{split}
\left<M\right>_{x',y'}&=2E_0\left[1-\frac{(\delta_1-\delta_2)^2}{8}+\frac{\delta_1+\delta_2}{2} \right]\cos(\theta) \\
& - \frac{E_0}{2}\sigma_{x'}^2\cos(\theta) - \frac{E_0}{2}\sigma_{y'}^2\frac{\cos(2\theta)}{\cos(\theta)}+O(3)\,.
\end{split}
\label{eq:InvariantMass-5}
\end{equation}

\section{Monochromatization with transverse correlation}
\begin{figure}[h]
\includegraphics[width=\columnwidth,trim=0 30 0 30, clip]{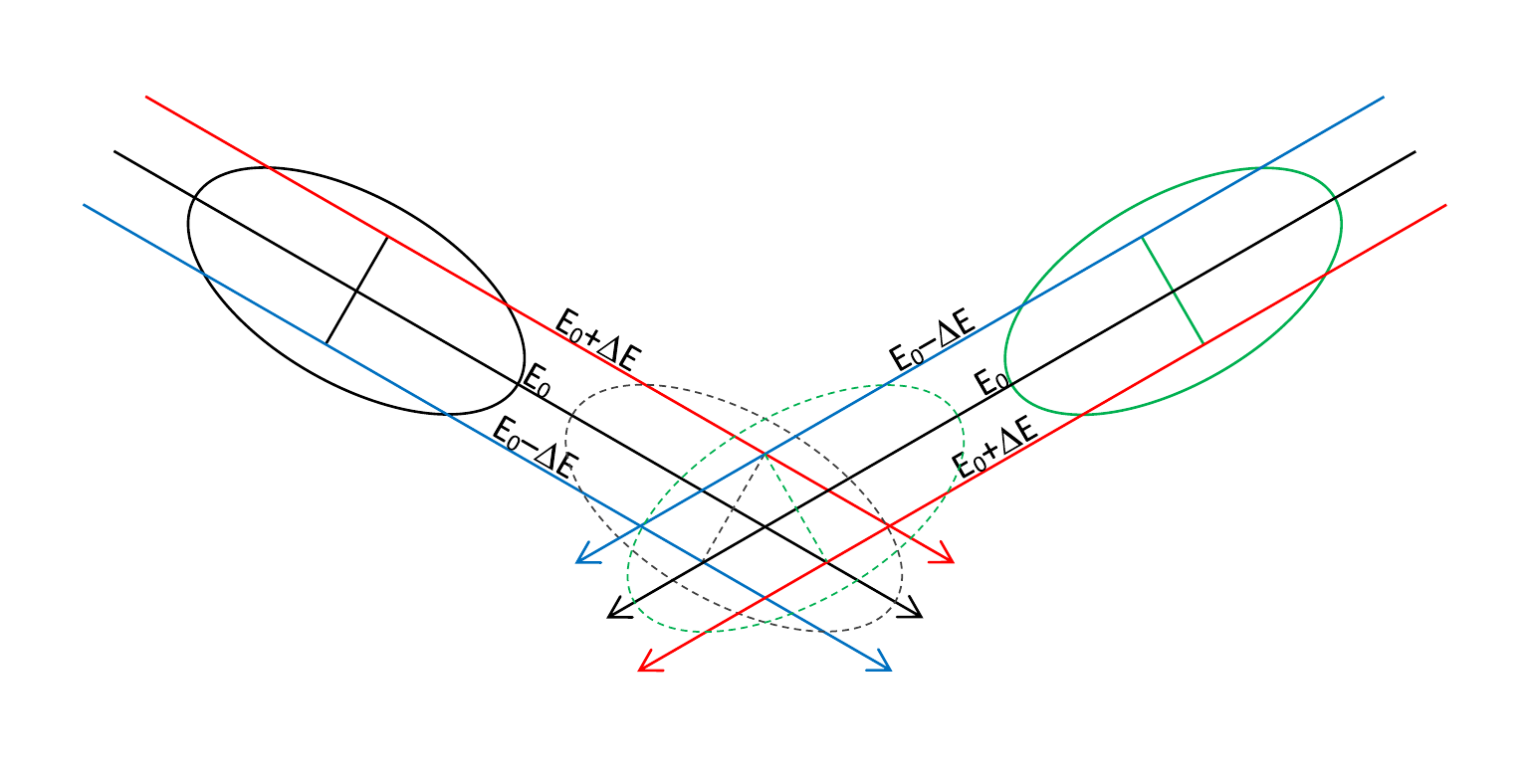}
\caption{Crossing angle collision with opposite dispersion.}
\label{fg:CrossingAngleCollisions-1}
\end{figure}

A.~Renieri \cite{ref:Renieri} proposed to introduce at IP horizontal dispersion $\pm\psi_x$ of opposite signs for colliding beams in head-on collisions. In this case the nature of monochromatization is obvious --- collision rate for particles with opposite energy deviation is higher than contrariwise (FIG. \ref{fg:Monochromatization-head-on}). Increasing collision angle (zero is head-on) will decrease efficiency of monochromatization but it is still possible (FIG. \ref{fg:CrossingAngleCollisions-1}) because particles with energy deviation will meet higher density of the particles with opposite energy deviation.

\begin{figure}[h]
\includegraphics[width=\columnwidth,trim=0 30 0 30, clip]{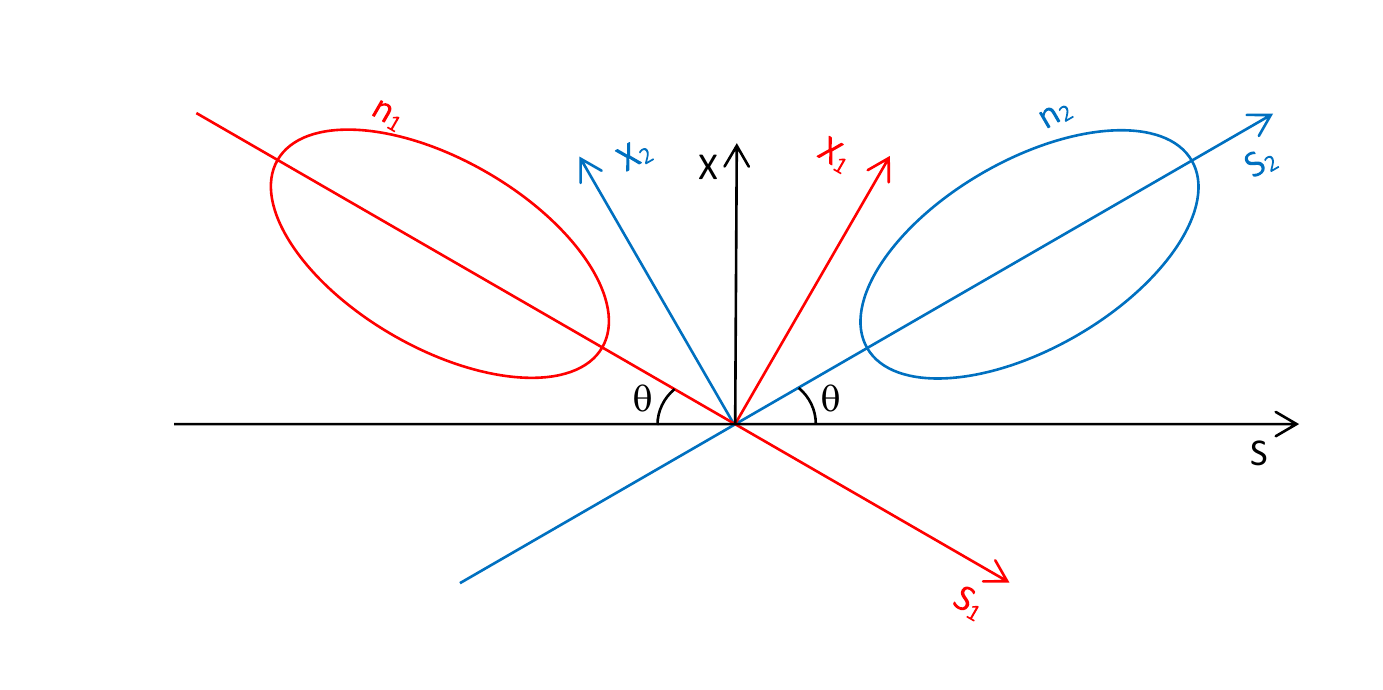}
\caption{Crossing angle collision with opposite dispersion.}
\label{fg:CrossingAngleCoordinates}
\end{figure}

For two colliding bunches (FIG. \ref{fg:CrossingAngleCoordinates}.)
we write the bunch density $n_i$ in own frame system
\begin{equation}
\label{eq:density-1}
\begin{split}
n_i(x_i,y_i,s_i,t,\delta_i)&=\frac{N_i}{(2\pi)^{4/2}\sigma_x\sigma_y\sigma_z\sigma_\delta}
\exp\!\left[
-\frac{(x_i\pm\psi_x\delta_i)^2}{2\sigma_x^2} \right.\\
&\quad \left.-\frac{(y_i\pm\psi_y\delta_i)^2}{2\sigma_y^2}-\frac{(s_i\mp ct)^2}{2\sigma_z^2}-\frac{\delta_i^2}{2\sigma_\delta^2}
 \right]\,,
\end{split}
\end{equation}
where we also introduced vertical dispersion $\pm\psi_y$ of opposite signs for colliding bunches, $N_i$ is bunch population, $\{x_i,y_i,s_i\}$ is usual accelerator basis of transverse and longitudinal coordinates, $\delta_i=(E_i-E_0)/E_0$ is energy deviation, and $\sigma_{x,y}=\sqrt{\varepsilon_{x,y}\beta_{x,y}}$, $\varepsilon_{x,y}$, $\beta_{x,y}$ are betatron beam size, emittance and beta function in corresponding plane at the IP, $\sigma_z$ and $\sigma_\delta$ are standard deviations of bunch longitudinal and energy deviation distributions. Then we need to transform coordinates into laboratory frame with the following expressions
\begin{equation}
\label{eq:transformation}
\left\{
\begin{aligned}
x_i&= x \cos(\pm\theta)+s \sin(\pm\theta)\,,\\
y_i&= y\,,\\
s_i&= x \sin(\pm\theta)+s \cos(\pm\theta)\,.
\end{aligned}
\right.
\end{equation}
Using (\ref{eq:luminosity-0}) we calculate luminosity according to
\begin{equation}
\label{eq:luminosity-1}
\begin{split}
\mathcal{L}&=f_0(1+\cos(2\theta))\times \\
&\times\int n_1(x,y,s,t,\delta_1)n_2(x,y,s,t,\delta_2)dx dy ds dct d\delta_1 d\delta_2\,.
\end{split}
\end{equation}
Neglecting hour glass effect the result is
\begin{align}
\label{eq:luminosity-2}
\mathcal{L}&=\frac{N_1N_2}{4\pi\sigma_x\sigma_y\sqrt{1+\varphi^2}}\frac{1}{\Lambda}=\frac{\mathcal{L}_0}{\Lambda}\,,\\
\label{eq:luminosity-3}
\frac{\partial^2\mathcal{L}}{\partial\delta_1\partial\delta_2}&=\frac{\mathcal{L}_0}{2\pi\sigma_\delta^2}
\exp\left[-\frac{(\delta_1-\delta_2)}{4\sigma_\delta^2}-\Lambda^2\frac{(\delta_1+\delta_2)}{4\sigma_\delta^2}\right]\,, \\
\label{eq:monochromatization-1}
\Lambda^2&=1+\frac{\psi_y^2\sigma_\delta^2}{\sigma_y^2}+\frac{\psi_x^2\sigma_\delta^2}{\sigma_x^2(1+\varphi^2)}\,,
\end{align}
where $\varphi=\sigma_z \tan{\theta}/\sigma_x$ is Piwinski angle, $\mathcal{L}_0$ is luminosity in absence of dispersion \cite{ref:Smith:1972}. Substitution of zero crossing angle $\theta$ and zero vertical dispersion will give expressions obtained by A.~Renieri.

Cross section of the narrow resonance with mass $M_0$ could be written as
\begin{equation}
\label{eq:CrossSection-1}
\sigma(\delta_1,\delta_2)=B+A\cdot\delta(\left<M\right>_{x',y'}-M_0)\,,
\end{equation}
where $B$ is describing background and independent of energy, $\left<M\right>_{x',y'}$ is given in \eqref{eq:InvariantMass-5}. With the help of \eqref{eq:luminosity-0} and \eqref{eq:luminosity-3} the production rate $\dot{N}$ in resonance vicinity is
\begin{equation}
\begin{aligned}
\dot{N}&=\int \sigma(\delta_1,\delta_2)\frac{\partial^2\mathcal{L}}{\partial\delta_1\partial\delta_2}d\delta_1d\delta_2 \\
&=B \frac{\mathcal{L}_0}{\Lambda}+A \frac{\mathcal{L}_0\exp\left[-\frac{m^2\Lambda^2}{4E_0^2\sigma_\delta^2}\right]}{\sqrt{2\pi}\cos(\theta)\sigma_\delta\sqrt{2E_0^2+E_0m\Lambda^2}}\,, \\
m&= \frac{E_0}{2}\sigma_{x'}^2 + \frac{E_0}{2}\sigma_{y'}^2\frac{\cos(2\theta)}{\cos(\theta)^2}+\frac{M_0}{\cos(\theta)}-2E_0\,.
\label{eq:Rate-1}
\end{aligned}
\end{equation}

Calculating expected values of \eqref{eq:InvariantMass-4} and \eqref{eq:InvariantMass-5} with \eqref{eq:luminosity-3} we obtain estimation of invariant mass resolution
\begin{equation}
\label{eq:InvariantMassResolution-1}
\begin{split}
\sigma_M^2&=\left<M^2\right>_{x',y',\delta_1,\delta_2}-\left<M\right>_{x',y',\delta_1,\delta_2}^2 \\
&=2E_0^2\left[\left(\frac{\sigma_\delta\cos(\theta)}{\Lambda}\right)^2+\left(\sigma_{x'}\sin(\theta)\right)^2\right]\,.
\end{split}
\end{equation}
Note that for large crossing angles the term with angular spread might be dominant.

\subsection{Notes on obtaining}
For a head-on collisions expression we may further simplify expression for $\Lambda$ \eqref{eq:monochromatization-1}. First, for $\Lambda\gg1$  we expand expression \eqref{eq:monochromatization-1}
\begin{equation}
\Lambda\approx \frac{\psi\sigma_\delta}{\sigma_x}\,.
\end{equation}
Second, we replace horizontal dispersion in the planar ring at the IP by expression \cite{ref:Courant:1997}
\begin{equation}
\label{eq:dispersion-1}
\begin{split}
\psi&=\frac{\sqrt{\beta_x}}{2\sin(\pi\nu_x)}\oint\frac{\sqrt{\beta_x(\tau)}}{\rho(\tau)}\cos(\varphi(\tau)-\pi\nu_x)d\tau \\
 &= \sqrt{\beta_x}F_D\,, \\
 F_D&=\frac{1}{2\sin(\pi\nu_x)}\oint\frac{\sqrt{\beta_x(\tau)}}{\rho(\tau)}\cos(\varphi(\tau)-\pi\nu_x)d\tau\,,
\end{split}
\end{equation}
where $\beta_x$ is horizontal beta function at the IP, $\rho$ is bending radius of the dipoles, $\varphi(\tau)$ is phase factor, $F_D$ is a factor defined by the whole lattice. 
Third, we substitute expressions for $\sigma_\delta$ and $\sigma_x=\sqrt{\varepsilon_x\beta_x}$ with the help of synchrotron radiation integrals \cite{ref:Helm:1973}
\begin{align}
\label{eq:sigma_delta}
\sigma_\delta&=\sqrt{C_q\gamma^2\frac{I_3}{J_sI_2}}\,,\\
\label{eq:sigma_x}
\sigma_x&=\sqrt{C_q\gamma^2\frac{I_5}{J_xI_2}\beta_x}\,, \\
\label{eq:I2}
I_2&=\oint\frac{d\tau}{\rho^2(\tau)}\,, \\ 
\label{eq:I3}
I_3&=\oint\frac{d\tau}{\left|\rho(\tau)\right|^3}\,, \\
\label{eq:I5}
I_5&=\oint\frac{\mathcal{H}(\tau)d\tau}{\left|\rho(\tau)\right|^3}\,,
\end{align}
where $J_x$ and $J_s$ are horizontal and longitudinal damping partition numbers, and  $\mathcal{H}(s)=\beta_x(s)\psi^{\prime 2}(s)+2\alpha_x(s)\psi(s)\psi^\prime(s)+\gamma_x(s)\psi^2(s)$.
Now then, we obtain expression for $\Lambda$
\begin{equation}
\Lambda\approx F_D\sqrt{\frac{J_x}{J_s}\frac{I_3}{I_5}}\approx F_D\sqrt{\frac{J_x}{J_s}\frac{1}{\left<\mathcal{H}\right>}}\,,
\end{equation}
where $\left<\right>$ denote average value over the ring.

So, in order to obtain large monochromatization factor $\Lambda$ it is necessary to have lattice with small horizontal emittance i.e. small $\left<\mathcal{H}\right>$, horizontal damping partition number greater than longitudinal  $J_x>J_s$, and large lattice factor $F_D$\eqref{eq:dispersion-1}.

\subsection{Example of FCC-ee}
Future circular collider (FCC) is a project in CERN of the next accelerator after LHC~\cite{ref25,ref26,ref:Oide:2016}. The ultimate goal is 100~km proton-proton machine with 100~TeV central mass energy. The first possible step is $e^+e^-$ factory --- FCC-ee with central mass energy range from 80~GeV to 350~GeV and two IPs. We apply derived expressions of transverse monochromatization scheme to see what luminosity and energy resolution could be achieved at Higgs boson threshold ($\sigma(ee\rightarrow H)=1.6$~fb, $\Gamma=4.2$~MeV, beam energy is $E_0=62.5$~GeV). 

Influence of synchrotron radiation of the particles in a strong electromagnetic field of the opposite bunch ({\it beamstrahlung}) modifies beam emittance, energy spread and length \cite{ref:Bogomyagkov:2013}. Presence of dispersion at the IP will change beam emittance even more; therefore calculations were performed in steps, on each step new parameters were calculated for a test bunch and then interchanged with parameters of the oncoming bunch until equilibrium is reached\cite{ref:Bogomyagkov:2013}. Table \ref{tbl:FCC} presents parameters at 62.5~GeV beam energy for baseline (no monochromatization), monochromatization with $\psi=1$~m and $\psi=0.5$~m.
\begin{table}[h]
\caption{Parameters of FCC-ee with and without monochromatization for two IPs. }
\label{tbl:FCC}
\begin{ruledtabular}
    \begin{tabular}{l|c|c|c}
												& Test 0				& Test 1					& Test 2					\\ \hline
Energy, GeV									& 62.5					& 62.5						& 62.5						\\ \hline
$\psi_x^*$, m								& 0					& 1						& 0.5						\\ \hline
$\beta_x^*/\beta_y^*$, mm				& 500/1				& 500/1					& 500/1.4					\\ \hline
$\varepsilon_x/\varepsilon_y$, nm/pm	& 0.26/2				& 2.8/21					& 3.8/2.9					\\ \hline
$\theta$, mrad								& \multicolumn{3}{c}{15}													\\ \hline
$\sigma_x$, mm								& 0.01					& 0.04						& 0.04						\\ \hline
$\sigma_y$, $\mu$m						& 0.04					& 0.15						& 0.2						\\ \hline
$\sigma_z$, mm								& 8.3					& 3.8						& 2.8						\\ \hline
$\sigma_{\delta}\,, 10^{-4}$				& 16 					& 7						& 8						\\ \hline
Piwinski angle $\varphi$					& 11					& 1.5						& 0.95						\\ \hline
$V_{RF}$, GeV								& 0.75					& 0.75						& 1.5						\\ \hline
$N_p$											& \multicolumn{3}{c}{$3\times 10^{11}$}								\\ \hline
$N_{bunches}$								& \multicolumn{3}{c}{2615}												\\ \hline
Luminosity,									& 						&							&							\\
$10^{34}$ cm$^{-2}$s$^{-1}$			& 9 					& 2.8						& 5					 	\\ \hline
$\Lambda$									& 1					& 10						& 6						\\ \hline
$\sigma_M$, MeV							& 141					& 6						& 11						\\
    \end{tabular}
\end{ruledtabular}
\end{table}
Notice with $\Lambda=10$ energy resolution improves more than by a factor of ten that is because of energy spread reduction. This happens owing to beamstrahlung: dispersion at the IP increases horizontal emittance \eqref{eq:sigma_x} and \eqref{eq:I5}, larger emittance means smaller beam density and weaker fields of the opposite bunch; therefore, smaller energy spread. In the last column (Test 2) we increased RF voltage and vertical beta function in order to match the length of interaction area with vertical beta function.

\section{Monochromatization with longitudinal correlation}
\begin{figure}[h]
\includegraphics[width=\columnwidth,trim=0 30 0 30, clip]{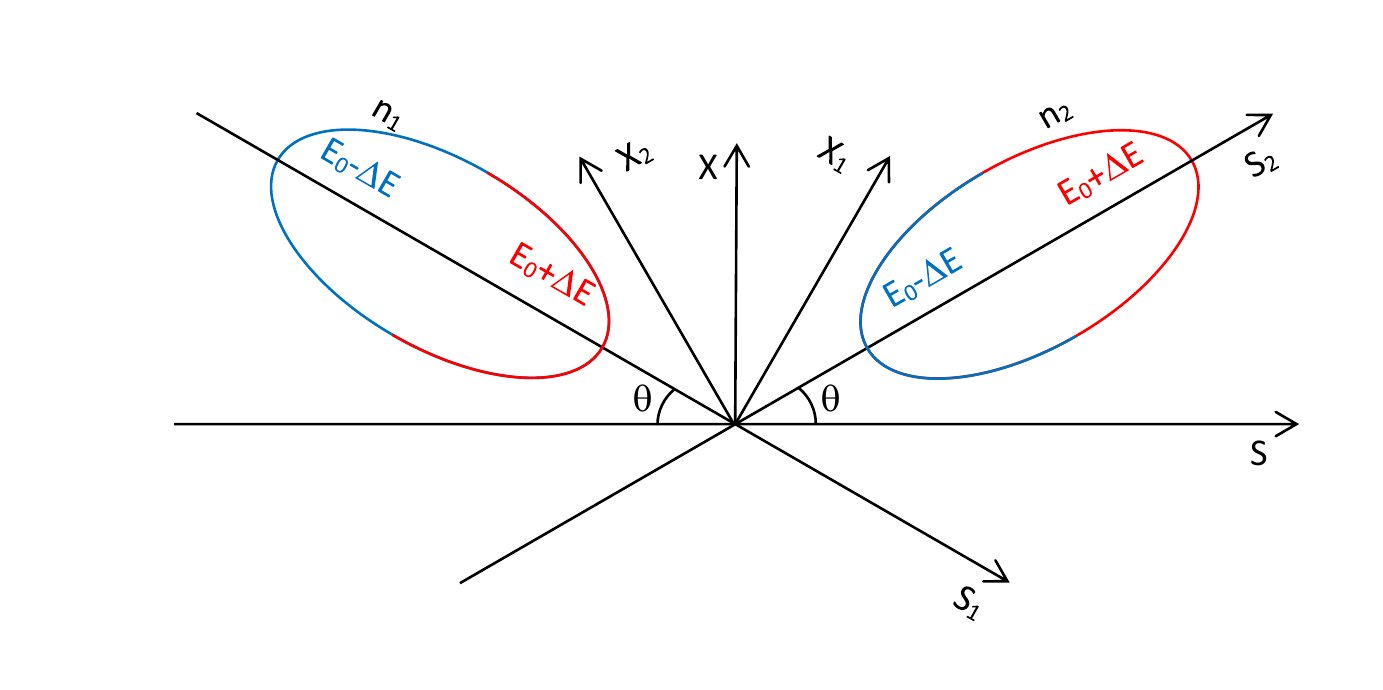}
\caption{Monochromatization for crossing angle collision with correlation between particle's energy and longitudinal position.}
\label{fg:CrossingAngleCollisions-2}
\end{figure}

Another approach to enchance collider energy resolution is to use correlation between particle's longitudinal position and energy (FIG. \ref{fg:CrossingAngleCollisions-2}).
Such longitudinal correlation may take place in case of strong focusing in longitudinal plain \cite{ref:Litvinenko:1996,ref:Gallo:2003:1,ref:Gallo:2003:2}. Figure 6 illustrates idea in the phase diagram, where particles belonging to two different ellipses collide when they reach the same $z$ but posses opposite sign of energy deviation.
\begin{figure}[h]
\includegraphics[width=\columnwidth,trim=0 10 0 10, clip]{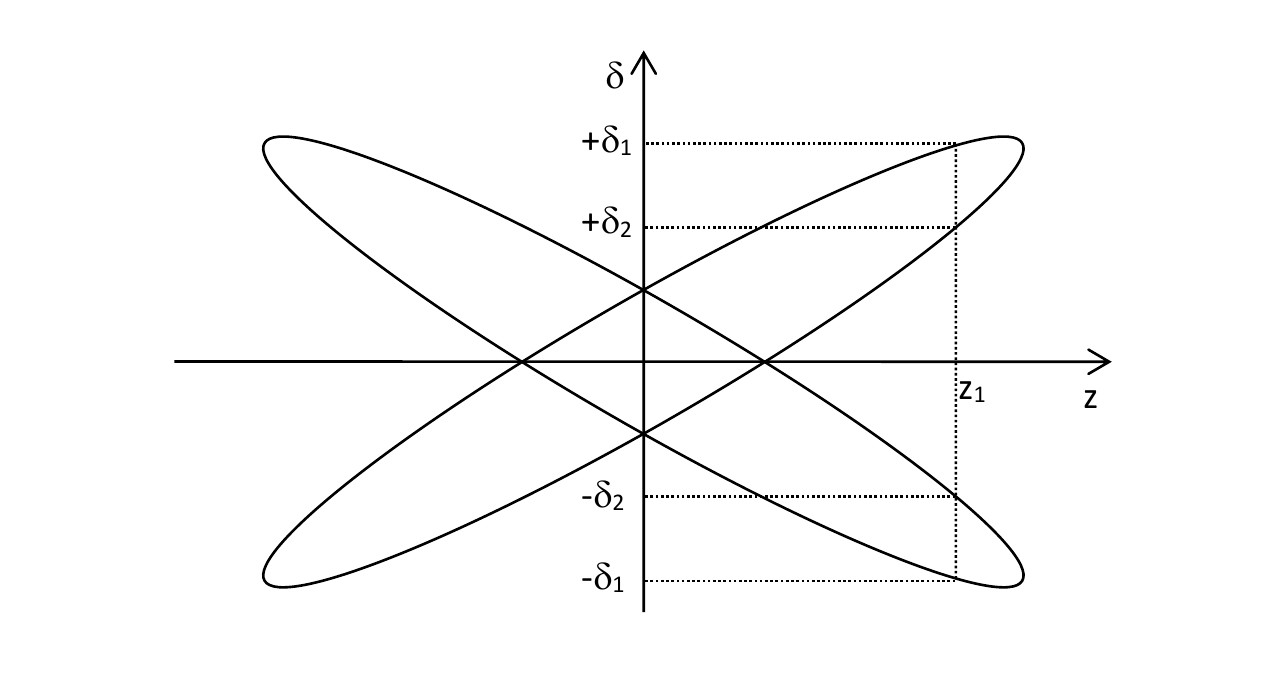}
\caption{Colliding beam ellipses in longitudinal plain, showing correlation between collision energy deviation $\delta$ and longitudinal position $z$.}
\label{fg:PhaseEllipse}
\end{figure}
Having $\beta_z, \alpha_z, \gamma_z, \varepsilon_z$ Twiss parameters and emittance in longitudinal plain with $\{z=s-ct,\delta\}$ conjugate variables, beam distribution in $\{z,\delta\}$ plain is  \cite{ref:Kapchinsky:1959}
\begin{equation}
\label{eq:density-2}
f(z,\delta)=\frac{1}{2\pi\varepsilon_z}\exp\left[-\frac{\gamma_z z^2+2\alpha_z z\delta+\beta_z\delta^2}{2\varepsilon_z}\right]\,.
\end{equation}
In order for monochromatization to happen, the bunches should posses opposite sign of $\alpha_z$ (FIG. \ref{fg:PhaseEllipse}), but because time $s$ flows in opposite direction for the bunches, in further calculations $\alpha_z$ has the same sign for both bunches.
The bunch density is
\begin{equation}
\label{eq:density-3}
\begin{split}
n_i(x_i,y_i,z_i,\delta_i)&=\frac{N_i}{(2\pi)^{4/2}\sigma_x\sigma_y\varepsilon_z}
\exp\!\left[
-\frac{x_i^2}{2\sigma_x^2} \right.\\
&\quad \left.-\frac{y_i^2}{2\sigma_y^2} -\frac{\gamma_z z^2+2\alpha_z z\delta+\beta_z\delta^2}{2\varepsilon_z}
 \right]\,,
\end{split}
\end{equation}
where $z_i=s_i\mp ct$. Calculations of luminosity by \eqref{eq:luminosity-1} with coordinate transformation \eqref{eq:transformation} gives similar to \eqref{eq:luminosity-2}, \eqref{eq:luminosity-3}, \eqref{eq:monochromatization-1} results but with different definition of $\Lambda$ and total luminosity is not decreased by $\Lambda$
\begin{align}
\label{eq:luminosity-4}
\mathcal{L}&=\frac{N_1N_2}{4\pi\sigma_x\sigma_y\sqrt{1+\varphi^2}}=\mathcal{L}_0\,,\\
\label{eq:luminosity-5}
\frac{\partial^2\mathcal{L}}{\partial\delta_1\partial\delta_2}&=\frac{\mathcal{L}_0 \Lambda}{2\pi\sigma_\delta^2}
\exp\left[-\frac{(\delta_1-\delta_2)}{4\sigma_\delta^2}-\Lambda^2\frac{(\delta_1+\delta_2)}{4\sigma_\delta^2}\right]\,, \\
\label{eq:monochromatization-2}
\Lambda^2&=\frac{1+\varphi^2}{1+\frac{1}{1+\alpha_z^2}\varphi^2}\approx 1+\alpha_z^2\,,\quad\varphi\gg1\,,
\end{align}
where $\sqrt{\varepsilon_z\beta_z}=\sigma_z$, $\sqrt{\varepsilon_z\gamma_z}=\sigma_\delta$, $\varphi=\sqrt{\varepsilon_z\beta_z}\tan(\theta)/\sigma_x$ is Piwinski angle. Note that in case of low $\alpha_z$, monochromatization vanishes $\Lambda=1$, which corresponds to the absence of the phase ellipses tilt (FIG. \ref{fg:PhaseEllipse}) and is a usual state of operation for modern colliders (bunch energy spread and length are independent of $s$). 

Production rate at the resonance vicinity with cross section \eqref{eq:CrossSection-1} is
\begin{equation}
\begin{aligned}
\dot{N}&=\int \sigma(\delta_1,\delta_2)\frac{\partial^2\mathcal{L}}{\partial\delta_1\partial\delta_2}d\delta_1d\delta_2 \\
&=B \mathcal{L}_0+A \frac{\Lambda\mathcal{L}_0\exp\left[-\frac{m^2\Lambda^2}{4E_0^2\sigma_\delta^2}\right]}{\sqrt{2\pi}\cos(\theta)\sigma_\delta\sqrt{2E_0^2+E_0m\Lambda^2}}\,, \\
m&= \frac{E_0}{2}\sigma_{x'}^2 + \frac{E_0}{2}\sigma_{y'}^2\frac{\cos(2\theta)}{\cos(\theta)^2}+\frac{M_0}{\cos(\theta)}-2E_0\,.
\label{eq:Rate-2}
\end{aligned}
\end{equation}
Comparing production rates for monochromatization with transverse correlation \eqref{eq:Rate-1} and longitudinal \eqref{eq:Rate-2} we notice that longitudinal monochromatization is contrary to transverse --- background rate is not reduced by $\Lambda$, and production rate of the resonance is increased by $\Lambda$. 

Invariant mass resolution is now $s$ dependent and independent of $\alpha_z$
\begin{equation}
\label{eq:InvariantMassResolution-2}
\begin{split}
\sigma_M^2&=\left<M^2\right>_{x',y',\delta_1,\delta_2}-\left<M\right>_{x',y',\delta_1,\delta_2}^2 \\
&=2E_0^2\left[\left(\frac{\sigma_\delta\cos(\theta)}{\Lambda}\right)^2+\left(\sigma_{x'}\sin(\theta)\right)^2\right] \\
&\approx 2E_0^2\left[\left(\sqrt{\frac{\varepsilon_z}{\beta_z(s)}}\cos(\theta)\right)^2+\left(\sigma_{x'}\sin(\theta)\right)^2\right]\,.
\end{split}
\end{equation}
In order to enhance invariant mass resolution one needs to decrease longitudinal emittance and increase beta function at the observation point.

\subsection{Examples of RF focusing}
Introduction of Twiss functions in longitudinal plane allowed authors of \cite{ref:Gallo:2003:1,ref:Chao:1979} to estimate longitudinal emittance
\begin{equation}
\label{eq:RFemittance}
\varepsilon_z=\sigma_{0\delta}^2\frac{\oint \frac{\beta(\tau)}{\left| \rho(\tau)\right|^3}d\tau}{\oint \frac{1}{\left| \rho(\tau)\right|^3}d\tau}\,,
\end{equation}
where $\sigma_{0\delta}^2=C_q\gamma^2I_3/(2I_2)$ is a usual expression for energy spread, $\rho(\tau)$ is bending radius. Obtained expression is similar and could obtained in the same manner as lower fundamental limit on vertical emittance.  For a planar isomagnetic ring with the help of \eqref{eq:InvariantMassResolution-2} we estimate invariant mass resolution 
\begin{equation}
\label{eq:InvariantMassResolution-3}
\sigma_M(s)\approx
\sqrt{2}E_0\sqrt{\sigma_{0\delta}^2\frac{\left<\beta_z\right>}{\beta_z(s)}\cos^2(\theta)+\sigma_{x'}^2\sin^2(\theta)}
\end{equation}
where $\left<\beta_z\right>$ is average value of longitudinal beta function in the ring. In order to enhance energy resolution we need to minimize $\left<\beta_z\right>$ and increase $\beta_z$, and minimize $\sigma_{x'}\sin(\theta)$.

\begin{figure}[h]
\includegraphics[width=\columnwidth,trim=0 0 0 0, clip]{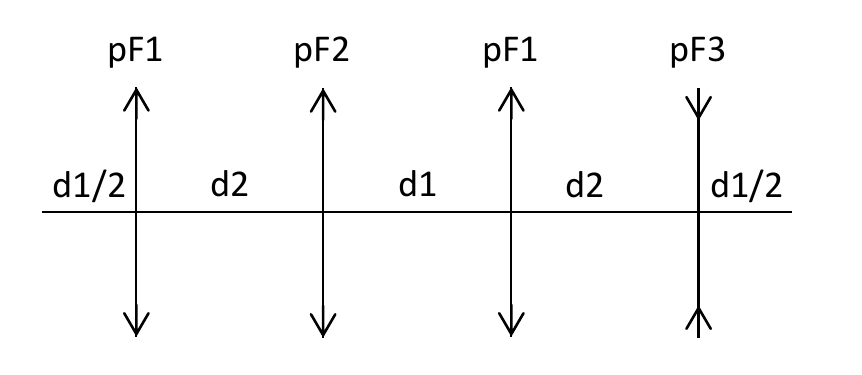}
\caption{Scetch of RF cavities of the toy ring.}
\label{fg:RFcavities}
\end{figure}

Authors of \cite{ref:Gallo:2003:1} studied the case of one cavity and reported a need for high voltage and short wave length RF. We decided to compare one cavity lattice with several cavities lattice. The latter is similar to using more complex cells than FODO in order to achieve lower emittance in transverse plain. For that reason we developed a toy lattice for $2\theta=90$ degree collision angle and with four RF cavities (FIG. \ref{fg:RFcavities}, TABLE \ref{tbl:MUMU}), where we chose $\sigma_{x'}=0$ to distinguish effect of RF focusing. Introduction of $\sigma_{x'}\neq 0$ will worsen invariant mass resolution.

The focusing strength of the cavity is \cite{ref:Litvinenko:1996}
\begin{equation}
\label{eq:FocusCavity}
\frac{1}{F}=\frac{V_{RF}}{E_0}\frac{2\pi}{\lambda_{RF}}\,,
\end{equation}
where $V_{RF}$ is RF voltage amplitude in the cavity, $\lambda_{RF}$ is wave length of RF field. Since RF amplitude is limited; therefore lower beam energy and shorter wave length are necessary to increase focusing strength of the cavity.
\begin{table}[h]
\caption{Parameters of the toy ring with longitudinal monochromatization.}
\label{tbl:MUMU}
\begin{ruledtabular}
\begin{tabular}{l|c|c|c}
												& Test 0				& Test 1					& Test 2					\\ \hline
Energy, MeV									& \multicolumn{3}{c}{150}												\\ \hline
$\Pi$, m										& \multicolumn{3}{c}{21}													\\ \hline
$2\theta$, rad								& \multicolumn{3}{c}{$\pi/2$}											\\ \hline
$\alpha$										& \multicolumn{3}{c}{0.5}													\\ \hline
$U0$, keV										& \multicolumn{3}{c}{2}													\\ \hline
$f0$, MHz										& \multicolumn{3}{c}{14.3}												\\ \hline
$q1/q2/q3$,									& \multicolumn{3}{c}{130/13/130}										\\ \hline
$\sigma_x$, $\mu$m						& \multicolumn{3}{c}{100}												\\ \hline
$\sigma_{x'}\,,10^{-3}$					& \multicolumn{3}{c}{0}													\\ \hline
$\sigma_{0\delta}\,,10^{-4}$				& \multicolumn{3}{c}{4}													\\ \hline
$V_{RF,1}/V_{RF,2}/V_{RF,3}$, MV		& 0/0.01/0			& 0/9/0					& -2.9/0.01/2.9			\\ \hline
$\varepsilon_z$, $10^{-6}$m				& 33					& 1						& 3.3						\\ \hline
$\sigma_{\delta,0}/\sigma_{\delta,1}/\sigma_{\delta,2}\,, 10^{-4}$		& 4/4/4 		& 4.9/4.9/4.9	& 5.3/5.3/5.3	\\ \hline
$\sigma_{z,0}/\sigma_{z,1}/\sigma_{z,2}$, mm	& 81/81/81	& 2.1/2.4/3.3				& 7/14/0.6				\\ \hline
$\sigma_{M,0}/\sigma_{M,1}/\sigma_{M,2}$, keV	& 60/60/60	& 71/63/46			& 70/35/800				\\
\end{tabular}
\end{ruledtabular}
\end{table}
We chose cavity $pF2$ with harmonic factor $q2$ as the one to compensate synchrotron radiation energy loss. The energy spread $\sigma_\delta$, bunch length $\sigma_z$ and invariant mass resolution $\sigma_M$ now depend on the azimuth, the index $\{0,1,2\}$ denotes positions at the beginning of the lattice, at the first and second cavities. Optimizing voltage and harmonic factors of other three cavities we achieved energy resolution almost twice better (Test 2) than in initial state (Test 0) and better than with one cavity (Test 1). The voltage of the additional cavities are 3~MV with frequency of 3~GHz lower than in one cavity case (Test 1). Higher beam energy will require higher voltage and shorter wave length, which now looks unrealistic. However, using more cavities and creating multi cell lattice with longitudinal final focus (like in transverse plain) might produce better monochromatization given that contribution from the term with $\sigma_{x'}$ is smaller.

\section{Conclusion}
We derived expressions for energy resolution and luminosity for crossing angle collisions with monochromatization based on transverse and longitudinal correlations between particle position and energy deviation. Monochromatization based on transverse correlation is feasible at the present level of accelerator technology; the longitudinal correlation requires high voltage and high frequency RF system, special longitudinal lattice --- hard to build at the present state of RF technology. Introduction of multi cavity lattice reduces requirements for RF system.

\begin{acknowledgments}
We express our gratitude to E.~Perevedentsev for reminding us about strong RF focusing.

We also wish to thank N.~Vinokurov for ingenious comment about dependence of invariant mass resolution on angular spread.

This work has been supported by Russian Science Foundation (project  N14-50-00080).

\end{acknowledgments}

\clearpage

\bibliography{References.bib}

\end{document}